\documentclass[a4paper,12pt,oneside,dvips,driver]{article}
\usepackage[english]{babel}
\usepackage{pstricks,pst-node,pst-coil}
\usepackage{epsfig}
\usepackage{amscd}
\usepackage{amssymb}
\usepackage[T1]{fontenc}
\usepackage{latexsym}
\usepackage[]{psfrag} 

\psfrag{el}{$\tilde{l}$}
\psfrag{eta0}{$\eta/(\sigma\xi_B)$}
\psfrag{tau1}{$\tau_1/\sigma\;$}
\psfrag{ggggtau2}{$\;\tau_2/\sigma$}
\psfrag{ft1t2ggh}{$\sigma f(\tau_1,\tau_2)\qquad $}
\psfrag{PGammay12}{$P_{\Gamma_1,\Gamma_2}$}
\psfrag{PGammay}{$P_{\Gamma}$}
\psfrag{y1}{$y_1$}
\psfrag{y2}{$y_2$}
\psfrag{GGamma1}{$G(\Gamma_1=0,\Gamma_2,L=100\xi_2)/\xi_2^2$}
\psfrag{Gamma2xi2}{$\Gamma_2/\xi_2$}

\newcommand{\beq}{\begin{equation}}
\newcommand{\eeq}{\end{equation}}
\newcommand{\bdm}{\begin{displaymath}}
\newcommand{\edm}{\end{displaymath}}

\begin{document}

\author{P. Jakubczyk, M. Napi\'{o}rkowski  \\
Instytut Fizyki Teoretycznej, Uniwersytet Warszawski \\ 00-681 Warszawa, Ho\.za  69, Poland   \\} 
\title{Interfacial morphology and correlations in adsorption at a chemically structured substrate - exact results in d=2} 
\date{} 
\maketitle 

\begin{abstract}
{Adsorption at a 1-dimensional planar substrate equipped with a localized chemical inhomogeneity is studied within the framework of a continuum interfacial model from the point of view of interfacial morphology and correlation function properties. Exact expressions for the one-point and two-point probability distribution functions $P_\Gamma (l_\Gamma)$ and $P_{\Gamma_1, \Gamma_2}(l_{\Gamma_1},l_{\Gamma_2})$, $l_\Gamma$ being the interface position above a fixed point $\Gamma$ of the substrate, are derived for temperature corresponding to the inhomogeneity's wetting transition. It is demonstrated that in the limit of macroscopic inhomogeneity's size the net effect of the remaining homogeneous parts of the substrate on the interfacial morphology above the inhomogeneity is exactly equivalent to appropriate pinning of the interface at its boundaries. The structure of the average interfacial morphology and correlation function in this limit are discussed and compared to earlier results obtained for systems with  homogeneous substrate.} \\


\noindent{\noindent}
\end{abstract}

\newpage

\section{Introduction}
\renewcommand{\theequation}{1.\arabic{equation}} 
\setcounter{equation}{0}
\vspace*{0.5cm}
Adsorption phenomena taking place at chemically patterned and geometrically sculptured substrates have been a topic of growing interest in recent years [1-8]. Theoretical research in this field has been stimulated on one side by rapid development of experimental methods of imprinting solid substrates with structures of sizes reaching nanometer scale [9-12], and on the other by attractive future applications, e.g. in the context of micro- and nanofluidics or nanotubes [13-14].

From theoretical point of view the interfacial morphologies and surface phase transitions at inhomogeneous substrates are most often studied via methods based on mean-field approach which correctly describe the system's behaviour provided intermolecular forces are long-ranged or the system's dimensionality is large enough. On the other hand, fluctuations often modify the system's properties, which is best seen in the context of 3d wetting with short-ranged forces \cite{3dW} or filling transitions in wedge-like geometries \cite{Parry1}. Additional interesting features emerge in the case of two-dimensional systems with chemically structured substrates, where interfacial fluctuations were shown to generate divergences in thermodynamic functions, in particular - the point tensions (see [16-17]). Exact results dealing with the case of inhomogeneous substrates are still rare though (see [16,18-19]). 
 
In this paper we are concerned with fluctuation effects accompanying adsorption at a one-dimensional planar substrate equipped with a single chemical impurity of width $2L$. The system under study consists of a two-dimensional fluid in a thermodynamic state infinitesimally close to bulk liquid-vapor coexistence and exposed to the aforementioned substrate, see Fig.1. The chemical structure imposes non-uniformity of the adsorbed liquid-like layer, whose morphology and fluctuations are investigated theoretically in the present paper.

\begin{figure}[h!]
\begin{center}
\includegraphics[width=0.75\textwidth]{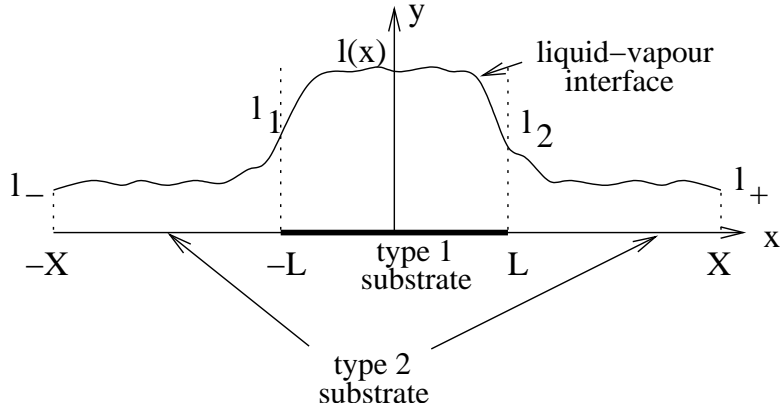}
\caption{A planar substrate equipped with a chemical inhomogeneity is exposed to fluid at bulk liquid-vapour coexistence. A layer of a liquid-like phase is adsorbed near the substrate. The interface at $y=l(x)$ separates the adsorbed liquid-like layer from the gas phase which is stable in the bulk. Type 1 substrate is wetted by the liquid, while substrate type 2 remains non-wet.} 
\end{center}
\end{figure}

The system's temperature is fixed at $T=T_{W1}$, the critical wetting temperature of homogeneous substrate type 1, of which the chemical heterogeneity is composed, see Fig.1. The rest of the substrate, which is referred to as type 2, remains non-wetted by the liquid; the corresponding wetting temperature $T_{W2}$ is above $T_{W1}$. The behaviour of the system in the limit of macroscopic inhomogeneity width $2L\to\infty$ is of our main interest here, since it corresponds to a typical experimental setup in which the heterogeneity's size is much larger than all correlation lengths present in the system. One also notes that the fluctuations' influence on the system is most pronounced in this limit. On phenomenological grounds, confirmed by earlier theoretical predictions [18,20] one may expect that the mean position of the  interface above the inhomogeneity centre, i.e. at $x=0$, as well as interfacial roughness and parallel interfacial correlation length diverge for $L\to\infty$. On the other hand, as was demonstrated in Ref.\cite{NapP}, these quantities remain bounded above the chemical impurity's boundaries at $x=\pm L$. It is then natural to ask about the behaviour of the characteristic local lengths in the intermediate region $|x|\in ]0,L[$, i.e., how strongly the fluctuations are suppressed by the non-wet parts of the substrate depending on the distance from the heterogeneity's boundary.    

Our approach is based on a continuum interfacial model and allows for exact derivation of the equilibrium probability distribution function $P_\Gamma (l_\Gamma)$ of finding the non-uniform two-phase interface at height $l_\Gamma$ above a fixed point $x=\Gamma$ located at the inhomogeneity, as well as the two-point distribution function $P_{\Gamma_1, \Gamma_2}(l_{\Gamma_1},l_{\Gamma_2})$.

The outline of this paper is as follows: in Section 2 we formulate the SOS model for the system under study and describe the methods we use to obtain our results. In Section 3 we derive and discuss the one-point probability distribution function $P_\Gamma (l_\Gamma)$ as well as the two-point probability distribution function $P_{\Gamma_1,\Gamma_2}(l_{\Gamma_1}, l_{\Gamma_2})$. Section 4 contains calculation of the average interfacial shape $\langle l_\Gamma\rangle$ as well as interfacial roughness $\langle l_\Gamma^2\rangle$ corresponding to the limit of macroscopic domain width. In Section 5 the two-point probability distribution function is utilized in order to obtain an expression for the interfacial correlation function, which depends on the two coordinates $\Gamma_1$, $\Gamma_2$. In Section 6 the results are summarized and compared to earlier predictions.

\section{The SOS model}
\renewcommand{\theequation}{2.\arabic{equation}} 
\setcounter{equation}{0}
\vspace*{0.5cm} 

The present study is based on a continuum interfacial SOS-type Hamiltonian model, within which one characterizes the system's state is by a single-valued, non-negative function $l(x)$ describing position of the interface above the substrate at location $x$. As was demonstrated in Ref.\cite{Burk} for the homogeneous substrate case, close to wetting temperature this approach captures all the system's properties relevant at length scales much larger than bulk correlation lengths, inter alia, it predicts the same set of surface critical indices as those obtained within 2D Ising model \cite{AbrahamI}. 

For the case of bulk fluid infinitesimally close to liquid-vapour coexistence, i.e. for $\mu=\mu_0^-$, $\mu_0$ being the chemical potential at coexistence, the SOS Hamiltonian has the following structure
\beq
\label{Hamiltonian}
\mathcal{H}[l]=\int_{-X}^X dx\Big[\frac{\sigma}{2}\Big(\frac{dl}{dx}\Big)^2+V(x,l)\Big]\; ,
\eeq
where $\sigma$ is the interfacial stiffness parameter, and $V(x,l)$ denotes the interfacial potential energy favouring the interface located near the substrate. In the present analysis $V(x,l)$ is taken as
\beq
\label{Potential}
 V(x,l)=\Theta (L-|x|)V_1(l)+\Theta (|x|-L) V_2(l)\; ,  
\eeq    
where $\Theta (x)$ is the Heaviside function.
In the case of periodic boundary conditions $l(-X)=l(X)$, the system's partition function represented by the path integral $Z(X,L)=\int\mathcal{D}l e^{-\mathcal{H}[l]}$ [17,21] can be expressed as
\beq
\label{Z}
 Z(X,L)=\int dl_- dl_1 dl_2 Z_2(l_1, l_-, X-L)Z_1 (l_2, l_1,2L)Z_2(l_-, l_2, X-L)\; ,
\eeq
where $l_-=l(\pm X)$, and $l_1=l(-L)$, $l_2=l(L)$ are positions of the interface at the inhomogeneity boundaries. The expression for the partition function $Z_i(y_1,y_2,\lambda)$ corresponding to homogeneous substrate of type $i=1,2$, of length $\lambda$, and with interface endpoints fixed at the boundaries at heights $y_1$, $y_2$, respectively, was derived by Burkhardt (see Ref.\cite{Burk}) for the case in which the potential $V_i(l)$ acts only at very short distances. In this paper we are concerned exclusively with this case. The partition function $Z(X,L)$ given by (\ref{Z}) was evaluated for arbitrary temperatures in Ref.\cite{NapP}, where it was then used to discuss the excess point free energy as function of temperature and the width $L$. The probability distribution functions and their moments, which can be obtained within this model for the special case $T=T_{W1}$ have not been discussed so far.

\section{Probability distribution functions}
\renewcommand{\theequation}{3.\arabic{equation}} 
\setcounter{equation}{0}
\vspace*{0.5cm}  

\subsection{One-point probability distribution function}
  
The one-point probability distribution function of finding the interface at height $l_\Gamma$ above a point $\Gamma$ of the substrate is given by 
\beq
P_\Gamma (l_\Gamma)=\langle \delta (l(x=\Gamma)-l_\Gamma)\rangle\; ,
\eeq
where $\langle ....\rangle$ denotes averaging with the Boltzmann weight $e^{-H[l]}$, with $H[l]$ given by Eq.(\ref{Hamiltonian}); the factor $(k_BT)^{-1}$ is included into the Hamiltonian. For $\Gamma\in ]-L,L[$ the above formula can be rewritten as 
\beq
\label{P}
P_\Gamma (l_\Gamma)=\frac{1}{Z(X,L)}\int dl_-\int dl_1\int dl_2Z_2(l_1,l_-,X-L)\times
\eeq
$$
\times Z_1(l_\Gamma,l_1,L+\Gamma)Z_1(l_2,l_\Gamma,L-\Gamma)Z_2(l_-,l_2,X-L)\; .
$$
The partition functions $Z_i(y_1, y_2, \lambda)$ in Eq.(\ref{P}) have the following spectral representations in terms of orthonormal sets of solutions $\{\psi_n^{(i)}\}$ to the corresponding Schr\"odinger equations with eigenvalues $E_n^{(i)}$ (see Ref.\cite{Burk}):
\beq
Z_i(y_1, y_2, \lambda)=\sum_n \psi_n^{(i)}(y_1)\psi_n^{(i)*}(y_2)e^{-E_n^{(i)}\lambda}\, .
\eeq
For temperatures below  $T_{W2}$ the spectrum $\{E_n^{(2)}\}$ corresponding to type 2 substrate contains a negative lowest eigenvalue $E_0^{(2)}$. It follows, that replacing $Z_2(l_1,l_-,X-L)$ and $Z_2(l_-,l_2,X-L)$ in Eq.(\ref{P}) by $\psi_0^{(2)}(l_1)\psi_0^{(2)*}(l_-)e^{-E_0^{(2)}(X-L)}$ and $\psi_0^{(2)}(l_-)\psi_0^{(2)*}(l_2)e^{-E_0^{(2)}(X-L)}$, respectively, is equivalent to neglecting terms of the order $e^{E_0^{(2)}(X-L)}$, which vanish in the limit of infinite substrate size $X\to\infty$. Bearing this in mind one can straightforwardly perform the integration over $l_-$ in Eq.(\ref{P}) to obtain
\beq
P_\Gamma (l_\Gamma)=\frac{\int dl_1\int dl_2\psi_0^{(2)}(l_1)\psi_0^{(2)*}(l_2)Z_1(l_\Gamma,l_1,L+\Gamma)Z_1(l_2,l_\Gamma,L-\Gamma)}{\int dl_1\int dl_2\psi_0^{(2)}(l_1)\psi_0^{(2)*}(l_2)Z_1(l_2,l_1,2L)}\; ,
\eeq 
where the limit $X\to\infty$ has been taken. The remaining integrals can be performed after inserting the specific forms of $\psi_0^{(2)}$ and $Z_1$ corresponding to $T=T_{W1}$ (see Ref.\cite{Burk}), namely
\beq
\psi_0^{(2)}(y)=\sqrt{-2\tau_2}e^{\tau_2 y}\;,
\eeq
\beq
\label{ZZ1}
Z_1(y_1,y_2,\lambda)=\sqrt{\frac{\sigma}{2\pi\lambda}}\Big[e^{-\sigma(y_2-y_1)^2/2\lambda}+e^{-\sigma(y_1+y_2)^2/2\lambda}\Big]\; ,
\eeq
where $\tau_2=-\sqrt{2\sigma/\xi_2}\sim (T-T_{W2})$.
 It is convenient to introduce the following dimensionless variables: $\lambda_0=\sqrt{2L/\xi_2}$,\hspace{3mm} $\lambda_+=\sqrt{(L+\Gamma)/\xi_2}$, \hspace{3mm} $\lambda_-=\sqrt{(L-\Gamma)/\xi_2}$, \hspace{3mm} $y=l_\Gamma/(2\xi_{2\perp})$, where $\xi_2$ and $\xi_{2\perp}$ are the parallel and perpendicular correlation lengths describing interfacial fluctuations above infinite and uniform substrate of type 2. The quantities $\xi_2$, $\xi_{2\perp}$ can be expressed in terms of the stiffness parameter $\sigma$ and the eigenvalue $E_0^{(2)}$ \cite{Burk}, and both diverge as $T$ approaches $T_{W2}$, which is accompanied by the eigenvalue $E_0^{(2)}$ increasing to zero. In terms of the above dimensionless parameters the distribution $P_\Gamma (l_\Gamma)$ takes the following form
\beq
\label{PGamma}
P_\Gamma (y)=e^{\lambda_0^2}\Big[e^{\lambda_0^2}\textrm{Erfc}(\lambda_0)+\frac{2}{\sqrt{\pi}}\lambda_0-2\lambda_0^2 e^{\lambda_0^2} \textrm{Erfc}(\lambda_0)  \Big]^{-1}\times
\eeq
\bdm
\times\Bigg\{\textrm{Erfc}\Big(-\frac{y}{\lambda_+}+\lambda_+\Big)\Bigg[e^{-4y}\textrm{Erfc}\Big(-\frac{y}{\lambda_-}+\lambda_-\Big)+\textrm{Erfc}\Big(\frac{y}{\lambda_-}+\lambda_-\Big)\Bigg]+
\edm
\bdm
+\textrm{Erfc}\Big(\frac{y}{\lambda_+}+\lambda_+\Big)\Bigg[e^{4y}\textrm{Erfc}\Big(\frac{y}{\lambda_-}+\lambda_-\Big)+\textrm{Erfc}\Big(-\frac{y}{\lambda_-}+\lambda_-\Big)\Bigg] \Bigg\}\; ,
\edm
where Erfc($x$) stands for the complementary error function
\beq
\label{Erfc}
\textrm{Erfc}(x)=1-\textrm{Erf}(x)=\frac{2}{\sqrt{\pi}}\int_x^\infty e^{-t^2}dt\; .
\eeq
The probability distribution function $P_\Gamma (l_\Gamma)$ is a decreasing function of $l_\Gamma$ at any point $\Gamma$, see Fig.2. Asymptotically this decay is of exponential type with characteristic decay length set by $\xi_{2\perp}$.  
\begin{figure}[h!]
\begin{center}
\includegraphics[width=0.65\textwidth]{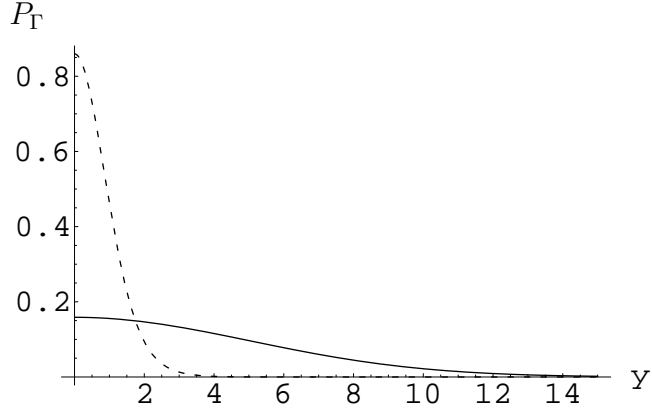}
\caption{The probability distribution function $P_\Gamma (y)$ plotted for $L/\xi_2=100$ and for two particular choices of $\Gamma$. The solid curve corresponds to the inhomogeneity center $\Gamma=0$, while the dashed one to the close vicinity of the inhomogeneity's boundary $\Gamma/L=0.99$. Approaching $\Gamma=0$ is accompanied by flattening the distribution function's profile, which indicates growth of the interfacial fluctuations' magnitude.} 
\end{center}
\end{figure}

\subsection{Two-point probability distribution function}

The two-point probability distribution function $P_{\Gamma_1, \Gamma_2}(l_{\Gamma_1},l_{\Gamma_2})$ of finding the interface at points $(\Gamma_1, l_{\Gamma_1})$ and $(\Gamma_2, l_{\Gamma_2})$ in the $(x,y)$ plane
\beq
P_{\Gamma_1, \Gamma_2}(l_{\Gamma_1},l_{\Gamma_2})=\langle \delta (l(x=\Gamma_1)-l_{\Gamma_1}) \delta (l(x=\Gamma_2)-l_{\Gamma_2})  \rangle\
\eeq
with $\Gamma_i\in (-L,L)$ can be written in the form 
\bdm
P_{\Gamma_1, \Gamma_2}(l_{\Gamma_1},l_{\Gamma_2})=
\edm
\beq
\frac{\int dl_1 \int dl_2\psi_0^{(2)}(l_1)\psi_0^{(2)*}(l_2)Z_1(l_{\Gamma 1},l_1,L+\Gamma_1)Z_1(l_{\Gamma_2},l_{\Gamma_1},\Gamma_2-\Gamma_1)Z_1(l_2,l_{\Gamma_2},L-\Gamma_2) }{\int dl_1\int dl_2\psi_0^{(2)}(l_1)\psi_0^{(2)*}(l_2)Z_1(l_2,l_1,2L)}\; ,
\eeq
where we assumed $\Gamma_2>\Gamma_1$ and passed to the limit $X\to\infty$.
After inserting the specific form of $Z_1$ corresponding to $T=T_{W1}$, Eq.(\ref{ZZ1}) into the above integrals one obtains
\beq
\label{P2Gamma}
P_{\Gamma_1, \Gamma_2}(y_1,y_2)=\Big[e^{\lambda_0^2}\textrm{Erfc}(\lambda_0)+\frac{2}{\sqrt{\pi}}\lambda_0-2\lambda_0^2 e^{\lambda_0^2} \textrm{Erfc}(\lambda_0) \Big]^{-1} \frac{1}{\sqrt{\pi\Delta\widetilde{\Gamma}}}   \times
\eeq
\bdm
\times e^{\lambda_{1+}^2+\lambda_{2-}^2}\Bigg(\textrm{Erfc}\Big(-\frac{y_1}{\lambda_{1+}}+\lambda_{1+}\Big)e^{-2y_1}+\textrm{Erfc}\Big(\frac{y_1}{\lambda_{1+}}+\lambda_{1+}\Big)e^{2y_1}\Bigg)\times
\edm
\bdm
\times \Bigg(\textrm{Erfc}\Big(-\frac{y_2}{\lambda_{2-}}+\lambda_{2-}\Big)e^{-2y_2}+\textrm{Erfc}\Big(\frac{y_2}{\lambda_{2-}}+\lambda_{2-}\Big)e^{2y_2}\Bigg)
\Bigg( e^{-\frac{(y_2-y_1)^2}{\Delta\widetilde{\Gamma}}}+e^{-\frac{(y_2+y_1)^2}{\Delta\widetilde{\Gamma}}}\Bigg)\; ,
\edm
where $\lambda_{1+}=\sqrt{(L+\Gamma_1)/\xi_2}$, \hspace{3mm} $\lambda_{2-}=\sqrt{(L-\Gamma_2)/\xi_2}$, \hspace{3mm} $y_1=l_{\Gamma_1}/(2\xi_{2\perp})$, \hspace{3mm} $y_2=l_{\Gamma_2}/(2\xi_{2\perp})$, \hspace{3mm} $\Delta\widetilde{\Gamma}=(\Gamma_2-\Gamma_1)/\xi_2$. A plot of $P_{\Gamma_1,\Gamma_2}(y_1,y_2)$ for a particular choice of system parameters is provided in Fig.3. 

\begin{figure}[h!]
\begin{center}
\includegraphics[width=0.85\textwidth]{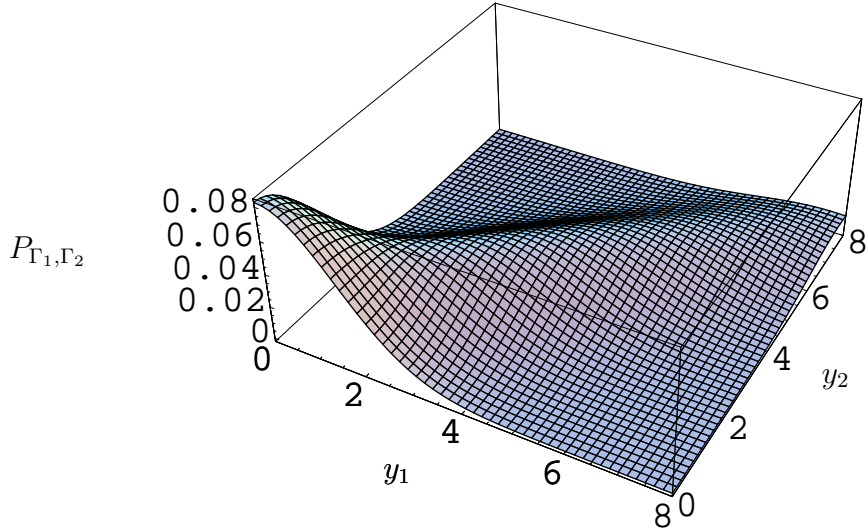}
\caption{The probability distribution function $P_{\Gamma_1,\Gamma_2} (y_1,y_2)$ plotted for $L/\xi_2=100$, $\Gamma_1/L=0.1$, $\Gamma_2/L=0.2$.} 
\end{center}
\end{figure}
In the next step we shall concentrate on deriving the moments $\langle y\rangle$ and $\langle y^2\rangle$ of the probability distribution function $P_\Gamma (y)$ as well as the moment $\langle y_1 y_2 \rangle$ of $P_{\Gamma_1, \Gamma_2}(y_1,y_2)$ in the asymptotic regime $L/\xi_2\gg 1$ and $(L-\Gamma)/\xi_2 \gg 1$. For this purpose we first note that the arguments of the error functions in $P_\Gamma (y)$ and $P_{\Gamma_1, \Gamma_2}(y_1,y_2)$ are of the order $\sqrt{L/\xi_2}$ (i.e., very large compared to 1) at any fixed $y$ (or $y_1$, $y_2$ in the case of $P_{\Gamma_1, \Gamma_2}(y_1,y_2)$). On the other hand for those values of $y$, $y_1$, $y_2$ for which the error functions' arguments become of the order unity, $P_\Gamma$ and $P_{\Gamma_1, \Gamma_2}$ are of the order $\sim e^{-L/\xi_{2\perp}}$, see Eq.(3.7), (3.11). It follows that the contributions to the integrals determining the aforementioned moments that are relevant in the large $L$ regime come only from those values of $y$, $y_1$, $y_2$, for which the arguments of the error functions are large. One may therefore utilize the asymptotic expansion of Erfc$(x)$ around infinity in order to compute the quantities of interest.

\section{Mean interfacial shape and roughness}
\renewcommand{\theequation}{4.\arabic{equation}} 
\setcounter{equation}{0}
\vspace*{0.5cm}
The mean equilibrium interfacial shape $\langle l_\Gamma\rangle$ is given by the first moment of the probability distribution function, Eq.(\ref{PGamma}), which we evaluate in the regime of macroscopic droplet height, i.e., for $L/\xi_2\gg 1$ and for $(L-\Gamma)/\xi_2 \gg 1$ by substituting Erfc$(\pm y/\lambda_{\pm}+\lambda_{\pm})$ with $\frac{1}{\sqrt{\pi}\lambda_{\pm}}e^{-(\pm y/\lambda_\pm+\lambda_{\pm})^2}$ . It follows that $\langle l_\Gamma\rangle$ fulfills the following relation 
\beq
\label{shape}
\frac{\langle l_\Gamma\rangle}{\sqrt{L}}\longrightarrow \frac{1}{\sqrt{\pi\sigma}}\sqrt{1-\frac{\Gamma^2}{L^2}} \;  
\eeq
for $L/\xi_2\to\infty$. The above expression corresponds to the upper root of an ellipse, whose long axis of length $L$ is oriented along the substrate, and the short axis' length equals $\sqrt{\frac{L}{\sigma\pi}}$. Note that the above asymptotic interfacial shape does not depend on $\tau_2$. We may now compare $\langle l_\Gamma\rangle$ in Eq.(\ref{shape}) to the droplet shape calculated for a system consisting of an interface fluctuating in presence of a homogeneous 1-dimensional substrate in the case where the interfacial endpoints are pinned (see Ref.\cite{Burk}). These shapes appear exactly the same, which means that the net influence the substrate at $x\in ]-\infty, -L[$ and $x\in ]L,\infty[$ exerts on the interfacial morphology above the chemical impurity is equivalent to pinning the interface at the inhomogeneity boundaries close to the substrate. This conclusion refers to the macroscopic limit $L\to\infty$. One also notes, that the scaling interfacial shape in the form of an ellipse was obtained for the case of long-ranged van der Waals type interactions via mean field approximation, see Ref.\cite{D2}. From Eq.(\ref{PGamma}) one finds the droplet height $\langle l_0\rangle$ divergent as $\sqrt{L}$ in the considered limit of large $L$. This may be compared to mean-field results [18,20] which predict logarithmic divergence of $\langle l_0\rangle$. As was argued in Ref.\cite{Conf}, the mean field theory gives correct prediction regarding $\langle l_0\rangle$ for d=3 system in which interfacial fluctuations are taken into account. From Eq.(\ref{shape}) one concludes, that for $\Gamma=\gamma L$, where $\gamma\in (0,1)$,  $\langle l_\Gamma\rangle$ diverges as $\frac{\sqrt{1-\gamma^2}}{\sqrt{\pi\sigma}}\sqrt{L}$, while in the case $\Gamma=L-\delta\Gamma$, $\delta\Gamma$ being small as compared to $L$, but much larger than $\xi_2$, one obtains $\langle l_\Gamma\rangle \sim\sqrt{\frac{2}{\pi\sigma}}(\delta\Gamma)^{1/2}$ for large $L$.    

The moment $\langle l_\Gamma^2\rangle$ can be computed for large $L$ analogously to $\langle l_\Gamma\rangle$. However, one may also derive this quantity without reference to any asymptotic expansion and for arbitrary values of $L$. This calculation is sketched in the Appendix. It appears, that for $L/\xi_2\gg 1$, $\langle l_\Gamma^2\rangle$ obeys the following scaling
\beq
\label{lG2}
\frac{\langle l_\Gamma^2\rangle}{L}\longrightarrow\frac{1}{2\sigma}\Big( 1-\frac{\Gamma^2}{L^2}\Big)\;.
\eeq
It follows, that the interfacial roughness $\langle l_\Gamma^2\rangle - \langle l_\Gamma\rangle^2$ diverges as $\frac{L}{\sigma}(\frac{1}{2}-\frac{1}{\pi})(1-\frac{\Gamma^2}{L^2})$, which is again the same formula, as one obtains by considering a homogeneous substrate with a fluctuating interface pinned at the boundaries.

\section{Interfacial correlation function}
\renewcommand{\theequation}{5.\arabic{equation}} 
\setcounter{equation}{0}
\vspace*{0.5cm}
The SOS model discussed in the present paper gives analytic insight into the structure of the correlation function in the limit $L/\xi_2\gg 1$. Let us recall that in the present analysis $T=T_{W1}$ which means that type 1 substrate is wetted by the liquid. This implies that in the homogeneous case the corresponding correlation length $\xi_1$ is infinite. On the other hand the fluctuations are suppressed by the presence of the outer, semi-infinite non-wet substrate parts. Whether for finite $L$ there exists a correlation length $\xi (L,\Gamma_1, \Gamma_2)$ describing the magnitude of correlations between interfacial fluctuations above points $\Gamma_1$ and $\Gamma_2$ is not obvious at all. Below we show, that no such quantity arises in the regime $L/\xi_2\gg 1$ and that the correlation function diverges linearly for $L/\xi_2\to\infty$ provided the points $\Gamma_1$, $\Gamma_2$ are chosen so that $(L-\Gamma_i)\gg \xi_2$ for $i=1,2$.   

The two-point interfacial correlation function $G(\Gamma_1,\Gamma_2,L)$ is defined by
\beq
G(\Gamma_1,\Gamma_2,L)=\langle l_{\Gamma_1} l_{\Gamma_2}\rangle-\langle l_{\Gamma_1}\rangle \langle l_{\Gamma_2} \rangle,
\eeq
where $\langle l_{\Gamma_1} l_{\Gamma_2}\rangle=\int dl_{\Gamma_1}dl_{\Gamma_2}  l_{\Gamma_1}l_{\Gamma_2} P_{\Gamma_1, \Gamma_2}(l_{\Gamma_1},l_{\Gamma_2})$. 

To evaluate $\langle l_{\Gamma_1} l_{\Gamma_2}\rangle$ we follow the scheme devised in the preceding sections and replace the error functions in $P_{\Gamma_1, \Gamma_2}(l_{\Gamma_1},l_{\Gamma_2})$ with their asymptotic at infinity. Integration over one of the variables is than straightforward, while the second integral is expressed in terms of the hypergeometric function $_2F_1$ \cite{Dilog} leading to
\beq
\label{cf1}
\langle y_1 y_2 \rangle = \frac{1}{M}\frac{4}{\pi^{3/2}}\frac{1}{\sqrt{\Delta\widetilde\Gamma}}\Big[\frac{\lambda_{1+}\lambda_{2-}\Delta\widetilde{\Gamma}^2}{2(\lambda_{1+}^2+\Delta\widetilde{\Gamma})(\lambda_{2-}^2+\Delta\widetilde{\Gamma})}+
\eeq
\bdm
+\frac{\lambda_{1+}^3\lambda_{2-}^3}{(\lambda_{1+}^2+\lambda_{2-}^2+\Delta\widetilde{\Gamma})^2} \,_2F_1\Big(\frac{1}{2},2,\frac{3}{2},-\frac{\lambda_{1+}^2\lambda_{2-}^2}{\Delta\widetilde{\Gamma}(\lambda_{1+}^2+\lambda_{2-}^2+\Delta\widetilde{\Gamma})}\Big)\Big]\;,
\edm
where
\beq
M=e^{\lambda_0^2}\textrm{Erfc}(\lambda_0)+\frac{2}{\sqrt{\pi}}\lambda_0-2\lambda_0^2e^{\lambda_0^2}\textrm{Erfc}(\lambda_0)\;.
\eeq
An example of the $G$ plot is provided in Fig.4.
\begin{figure}[h!]
\begin{center}
\includegraphics[width=0.65\textwidth]{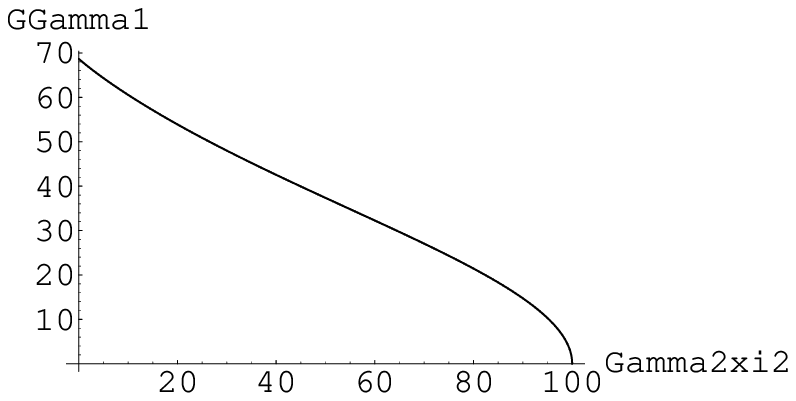}
\caption{The correlation function $G$ plotted as a function of $\Gamma_2$ at fixed $\Gamma_1=0$ and $L/\xi_2=100$ in the range $\Gamma_2\in (0,L)$.} 
\end{center}
\end{figure}
To obtain a more transparent expression for the correlation function $G(\Gamma_1,\Gamma_2,L)$ we focus on the case where the points $\Gamma_1$, $\Gamma_2$ are separated by a distance much shorter than $L$ and $L-\Gamma_2$, i.e., we assume $\Delta\widetilde{\Gamma}\ll\lambda_{1+}^2$ and $\Delta\widetilde{\Gamma}\ll\lambda_{2-}^2$. Using the asymptotic properties of the hypergeomertic function $_2 F_1$
\beq
_2 F _1 (\frac{1}{2},2,\frac{3}{2},-x)=\frac{\pi}{4}\frac{1}{\sqrt{\pi}}-\frac{1}{3}\frac{1}{x^2}+\mathcal{O} (\frac{1}{x^3})\;,
\eeq
we obtain the asymptotic behaviour of $G(\Gamma_1,\Gamma_2,L)$ in the form
\beq
\label{cf2}
\frac{G(\Gamma_1,\Gamma_2,L)}{L}\longrightarrow\frac{1}{\sigma}\Big[\frac{1}{2}\Big(1+\frac{\Gamma_1}{L}\Big)\Big(1-\frac{\Gamma_2}{L}\Big)-\frac{1}{\pi}\sqrt{1-\frac{\Gamma_1^2}{L^2}}\sqrt{1-\frac{\Gamma_2^2}{L^2}}\Big]\;,
\eeq
which shows that under the restrictions imposed on the parameters $\Gamma_1$, $\Gamma_2$ the decay of correlations is in fact linear in $\Delta\widetilde{\Gamma}$, and that for fixed $\Gamma_1$, $\Gamma_2$ the correlations grow linearly as a function of $L$. For $\Gamma_1=\Gamma_2$ one recovers the local roughness derived in Section 4, Eq.(4.2). The applicability of Eq.(\ref{cf2}) is strongly limited by the aforementioned restrictions. On the other hand also in formula (\ref{cf1}) one does not observe the presence of terms of type $e^{-\Delta\Gamma/\xi}$, which typically describe the decay of correlations in non-critical systems.

\section{Summary}
\renewcommand{\theequation}{6.\arabic{equation}} 
\setcounter{equation}{0}
\vspace*{0.5cm}
In this paper we analyzed the properties of a two-dimensional model of adsorption at a substrate equipped with a single chemical impurity. We were concerned with the case, when the inhomogeneity is wetted by the liquid phase of the fluid to which the substrate is exposed, while the rest of the substrate remains non-wet. The behaviour of mean local interfacial position, interfacial roughness and correlation function was investigated in the asymptotic regime of large inhomogeneity's width $2L$, for which the system becomes critical and is fluctuation-dominated. The type of divergence of these quantities in the considered limit was determined depending on the distance from the inhomogeneity's boundary. The results were obtained by directly calculating the one- and two-point probability distribution functions $P_\Gamma (l_\Gamma)$ and $P_{\Gamma_1, \Gamma_2}(l_{\Gamma_1},l_{\Gamma_2})$, and then investigating the asymptotic properties of their moments. Our conclusions are as follows:
\begin{itemize}
\item
Both mean interfacial position $\langle l_\Gamma\rangle$  and roughness $(\langle l_\Gamma^2\rangle-\langle l_\Gamma\rangle^2)^{1/2}$ diverge in the limit $L\to\infty$ at any point $\Gamma$ which is separated from the inhomogeneity's boundary by a distance much larger than the parallel correlation length $\xi_2$ corresponding to homogeneous substrate of type 2. For $\Gamma=\gamma L$, where $\gamma\in (0,1)$, the quantity $\langle l_\Gamma\rangle$ diverges as $\sqrt{1-\gamma^2}\sqrt{L}$, while for $\Gamma=L-\delta\Gamma$, $\delta\Gamma$ being small as compared to $L$, but much larger than $\xi_2$, one obtains $\langle l_\Gamma\rangle \sim (\delta\Gamma)^{1/2}$. The mean local height and roughness close to the chemical structure's boundaries remain bounded even for infinite $L$, as was already noted in Ref.\cite{NapP}. The character of the divergence of the local roughness is the same as for $\langle l_\Gamma\rangle$. 
\item
The limiting forms of the mean interfacial shape as well as interfacial roughness are independent of the properties of substrate 2 and are the same as those obtained for a system composed of an interface fluctuating above a homogeneous substrate of fixed length $2L$ provided the interface endpoints at $-L$ and $L$ are pinned at a finite distance from the substrate. This means that from the point of view of the system's large-scale behaviour in the region above the chemical structure, the presence of the non-wetted parts of the substrate can be exactly imitated by a system with a homogeneous substrate upon binding the interface at $-L$ and $L$.
\item
The obtained profiles of $\langle l_\Gamma\rangle$  and $(\langle l_\Gamma^2\rangle-\langle l_\Gamma\rangle^2)^{1/2}$ describe ellipses, whose $x$-axis length equals $2L$, and $y$-axis lengths are or the order $\sqrt{L}$ and are inversely proportional to $\sqrt{\sigma}$. 
\item
We investigated the structure of the interfacial correlation function $G(\Gamma_1,\Gamma_2,L)$ in the asymptotic limit of large $L$ and found no indication of presence of terms of type $e^{-(\Gamma_2-\Gamma_1)/\xi}$ ($\Gamma_2>\Gamma_1$), which often describe decay of fluctuations in non-critical systems. The correlation function's amplitude grows linearly with $L$, and the correlations decay linearly with $\Gamma_2-\Gamma_1$ under the restrictions that $\Gamma_2-\Gamma_1\ll L$ and $\Gamma_2-\Gamma_1\ll L-\Gamma_2 $.

\end{itemize}

\section{Appendix}
\renewcommand{\theequation}{A.\arabic{equation}} 
\setcounter{equation}{0}
\vspace*{0.5cm}
In this Appendix we discuss the expression for the moment $\langle l_\Gamma ^2 \rangle$ of the probability distribution function $P_\Gamma (y)$ given by Eq.(\ref{PGamma}). The quantity $\langle l_\Gamma ^2 \rangle$ is calculated for arbitrary $L$ and $\Gamma\in (-L,L)$ by noting that $P_\Gamma (y)$ is invariant with respect to the inversion $y\longrightarrow -y$. This allows us to write $\langle l_\Gamma ^2 \rangle$ in the following form
\beq
\langle y^2 \rangle=\int_{-\infty}^{\infty} dy y^2 \Big[\textrm{Erfc}(\lambda_0)+\frac{2}{\sqrt{\pi}}\lambda_0e^{-\lambda_0^2}-2\lambda_0^2 \textrm{Erfc}(\lambda_0)  \Big]^{-1}\times
\eeq
\bdm
\times \Bigg[e^{-4y}\textrm{Erfc}\Big(-\frac{y}{\lambda_+}+\lambda_+\Big)\textrm{Erfc}\Big(-\frac{y}{\lambda_-}+\lambda_-\Big)+\textrm{Erfc}\Big(-\frac{y}{\lambda_+}+\lambda_+\Big)\textrm{Erfc}\Big(\frac{y}{\lambda_-}+\lambda_-\Big)\Bigg]\; .
\edm
The above integrals are performed by substituting $U(y)=\textrm{Erfc}\Big(-\frac{y}{\lambda_+}+\lambda_+\Big)$, $V_1'(y)=y^2e^{-4y}\textrm{Erfc}\Big(-\frac{y}{\lambda_-}+\lambda_-\Big)$, $V_2'(y)=y^2\textrm{Erfc}\Big(\frac{y}{\lambda_-}+\lambda_-\Big)$, and integrating by parts. This way one arrives at an integral that no longer involves products of the error functions. In the next step we insert the integral representation (Eq.(\ref{Erfc})) of the error function into the obtained formula and integrate over $y$ and the remaining variable. This yields the following expression
\beq
\label{lG2S}
\langle y^2 \rangle=\frac{2}{\sqrt{\pi}} \Big[\textrm{Erfc}(\lambda_0)+\frac{2}{\sqrt{\pi}}\lambda_0e^{-\lambda_0^2}-2\lambda_0^2 \textrm{Erfc}(\lambda_0)  \Big]^{-1}\times
\eeq
\bdm
\times \Big[\Big(\frac{1}{12}\lambda_0^6+\frac{1}{12}\lambda_0^4+\frac{1}{8}\lambda_0^2+\lambda_0^2\widetilde{\Gamma}^2\Big)\frac{e^{-\lambda_0^2}}{\lambda_0} + \sqrt{\pi}\Big(-\frac{1}{12}\lambda_0^6-\frac{1}{8}\lambda_0^4-\lambda_0^2\widetilde{\Gamma}^2-\frac{1}{2}\widetilde{\Gamma}^2+\frac{1}{16}\Big)\textrm{Erfc}(\lambda_0)\Big]\;,
\edm
from which it follows, that $\langle y^2 \rangle$ is quadratic in $\widetilde{\Gamma}$ at arbitrary $L$. 

Utilizing the asymptotic expansion Erfc$(x)=\frac{1}{\sqrt{\pi} x}e^{-x^2}(1-\frac{1}{2x^2}+\frac{3}{4x^4}-\frac{15}{8x^6}+\mathcal{O}(\frac{1}{x^8}))$ one obtains the following formula valid for $L/\xi_2\gg 1$
\beq
\langle l_\Gamma^2 \rangle=\frac{L}{2\sigma}\Big(1-\frac{\Gamma^2}{L^2}\Big)+\xi_{2\perp}\Big(1+\frac{\Gamma^2}{L^2}\Big)+\mathcal{O}\Big(\frac{\xi_2}{L}\Big)\; ,
\eeq
which up to the leading terms is equivalent to Eq.(\ref{lG2}). 

One may also explore the asymptotic regime of Eq.(\ref{lG2S}) corresponding to $L/\xi_2\ll 1$. For this case we obtain
\beq
\langle l_\Gamma^2 \rangle=\frac{1}{2}\xi_{2\perp}\Big[1+2\lambda_0^2-\frac{8}{3\sqrt{\pi}}\lambda_0^3-8\Big(\frac{1}{4}\lambda_0^4+\widetilde{\Gamma}^2+\mathcal{O}(\lambda_0^{5})\Big)\Big]\;.
\eeq
For $L/\xi_2=0$ the result for homogeneous substrate type 2 is recovered. One observes that the two leading $L$-dependent corrections are not influenced by the value of $\Gamma$. 

The method we used to obtain $\langle l_\Gamma^2 \rangle$ for arbitrary $L$ may also be applied to compute higher even moments of the distribution $P_\Gamma (l_\Gamma)$. However, it fails in the case of the odd moments, which we were able to obtain only in the asymptotic regime $L/\xi_2\gg 1$, as described in the main text. 
 
\vspace{1cm}

\noindent {\bf {Acknowledgment}}  \\ 
\noindent This work has been supported by the  Ministry of Science and Higher
Education via the grant N202 076 31/0108.

\end{document}